\documentstyle[twoside,fleqn,espcrc2,epsf]{article}

\newcommand{\AmS}{{\protect\the\textfont2
  A\kern-.1667em\lower.5ex\hbox{M}\kern-.125emS}}

\newlength{\numlen}

\newcommand{\h}{{\hspace{0.5 cm}}}

\newcommand{\tr}{{\rm Tr\,}}
\newcommand{\nr}[1]{(\ref{#1})}
\newcommand{\fr}[2]{{\frac{#1}{#2}}}

\newcommand{\la}[1]{\label{#1}}

\newcommand{\figtopspace}{\vspace*{-0.5cm}}
\newcommand{\figbottomspace}{\vspace*{-4.5cm}}
\newcommand{\be}{\begin{equation}}
\newcommand{\ee}{\end{equation}}
\newcommand{\ba}{\begin{eqnarray}}
\newcommand{\ea}{\end{eqnarray}}
\newcommand{\bi}{\begin{itemize}}
\newcommand{\ei}{\end{itemize}}
\newcommand{\rmi}[1]{{\mbox{\scriptsize #1}}}

\newcommand{\nn}{\nonumber}
\newcommand{\lambdamsbar}{\Lambda_{\overline{\rm MS}}}
\newcommand{\dr}{{\rm dr}}
\newcommand{\seq}{\hspace*{-2.5mm}&=&\hspace*{-2.5mm}}

\def\lsi{\raise0.3ex
\hbox{$<$\kern-0.75em\raise-1.1ex\hbox{$\sim$}}}
\def\gsi{\raise0.3ex
\hbox{$>$\kern-0.75em\raise-1.1ex\hbox{$\sim$}}}

\newcommand{\gsim}{\mathop{\gsi}}

\newcommand{\eq}{eq.\,}

\newcommand{\figfxsize}{8cm}

\title{High-$T$ QCD and dimensional reduction: measuring
the Debye mass\thanks{Presented by K. Rummukainen at Lattice '97}}

\author{%
K. Kajantie\address{Theory Division, CERN, CH-1211 Geneva 23, Switzerland}%
$^,$\address{Department of Physics, P.O.Box 9,
00014 University of Helsinki, Finland},
M. Laine\address{Institut f\"ur Theoretische Physik,
Philosophenweg 16, D-69120 Heidelberg, Germany},
J. Peisa\address{Department of Mathematical Sciences,
University of Liverpool, Liverpool L69 3BX, UK},
A. Rajantie$^{\rm b}$,
K. Rummukainen\address{Fakult\"at f\"ur Physik, Postfach 100131, D-33501
Bielefeld, Germany} and
M. Shaposhnikov$^{\rm a}$
\hfill\raisebox{22mm}[0mm][0mm]{\makebox[0mm][r]{\large BI-TP 97/35}}%
\raisebox{17mm}[0mm][0mm]{\makebox[0mm][r]{\large September 1997}}
}
\begin{document}

\begin{abstract}
We study the high-temperature phase of SU(2) and SU(3) QCD using
lattice simulations of an effective 3-dimensional SU($N$) + adjoint Higgs
-theory, obtained through dimensional reduction.  We investigate the
phase diagram of the 3D theory, and find that the high-$T$ QCD
phase corresponds to the metastable symmetric phase of the 3D theory.  We
measure the Debye screening mass $m_D$ with gauge invariant operators;
in particular we determine the ${\cal O}(g^2)$ and ${\cal O}(g^3)$
corrections to $m_D$.  The corrections are seen to be large, modifying
the standard power-counting hierarchy in high temperature QCD.
\vspace*{-3mm}
\end{abstract}

\maketitle
\thispagestyle{empty}



The Debye mass (inverse screening length of color electric fields)
characterizes the coherent static interactions in QCD plasma,
and its numerical value is essential for phenomenological discussion
of the physics of the QCD plasma.
For SU($N$) QCD, $N=2,3$, with $N_f$ massless 
quarks, the Debye mass can be expanded at high temperatures in a power series
in the coupling $g$:
\ba
  m_D & = & m_D^\rmi{LO}+{Ng^2T\over4\pi}\ln{m_D^\rmi{LO}\over  
  g^2T}\nonumber\\
  & + & 
  c_N g^2T + d_{N,N_f} g^3 T + {\cal O}(g^4T),
\label{md4d}
\ea
where the leading order perturbative result is $m_D^\rmi{LO}
=(N/3+N_f/6)^{1/2}gT$.  The logarithmic part of the ${\cal O}(g^2)$
correction can be extracted perturbatively \cite{rebhan}, but $c_N$
and the higher order corrections are non-perturbative.  Our aim is to
evaluate numerically the coefficients $c_N$ and $d_{N,N_f}$.  A
detailed discussion of the results can be found in
\cite{adjoint2,letter}.

An effective 3D action, obtained through {\em dimensional reduction\,}
\cite{old,generic,effbn}, is a powerful tool for studying
high-$T$ QCD.  The effective action can be derived perturbatively
without the infrared problems associated with the standard high-$T$
perturbative analysis.  It retains the essential infrared physics of
the original theory, and since it is bosonic even for $N_f>0$, it can
be studied economically with lattice Monte Carlo simulations.
Recently it has been very successfully applied to the Electroweak
phase transition \cite{ewreview}.

The effective action is derived with the Green's
function matching technique \cite{generic,effbn}:
\ba
L_\rmi{eff} & = &  {\mbox{$\fr14$}}  F_{ij}^aF_{ij}^a
+ \tr [D_i,A_0][D_i,A_0]  \nn \\
& + & m_3^2\tr A_0^2 +\lambda_A(\tr A_0^2)^2\,.
\la{leff}
\ea
$A_0$ is a remnant of the temporal gauge fields and belongs to the
adjoint representation of SU($N$).  The Lagrangian \nr{leff} gives the
Green's functions to a relative accuracy ${\cal
O}(g^4)$~\cite{generic}, sufficient for the accuracy of the expansion
in \eq\nr{md4d}.

For $N_f=0$, the 3D couplings $g_3^2$, $m_3^2$ and $\lambda_A$
are related to the temperature $T$ (and the 4D gauge coupling
$g^2(\mu)$, which is evaluated at the optimized scale~\cite{adjoint2}
$\mu \approx 2\pi T$) by
\ba
g_3^2 = g^2 T 
  \seq \fr{24\pi^2 T}{22 \log(6.742 T/\lambdamsbar)} \la{g3}\\
x \equiv \fr{\lambda_A}{g_3^2} 
  \seq \fr{6+N}{11N} \fr1{\log(5.371T/\lambdamsbar)}   \la{x} \\
y \equiv \frac{m_3^2}{g_3^4}   
  \seq \fr{2}{9\pi^2x}+\fr{4}{16\pi^2} +{\cal O}(x)
  \h\mbox{SU(2)} \la{y2} \\
  \seq \fr{3}{8\pi^2x}+\fr{9}{16\pi^2}+{\cal O}(x)
  \h\mbox{SU(3)} \la{y3}
\ea
The dynamics of the 3D theory is fully characterized by the
dimensionless ratios $x$ and $y$ above and the dimensionful gauge
coupling $g_3^2$.  The presence of fermions only modifies the
numerical factors in eqs.\,(\ref{g3}--\ref{y3}).

Due to the {\em superrenormalizability} of the 3D action the
continuum$\leftrightarrow$lattice relations of the couplings become
very transparent (for a detailed discussion, see
\cite{adjoint2,contlatt}).  In particular, the lattice gauge coupling
$\beta_G$ is related to the continuum gauge coupling $g_3^2$ and the lattice
spacing $a$ by $\beta_G = 2N/(g_3^2\,a)$.

\begin{figure}[b]
\vspace*{-3mm}
\figtopspace
\epsfxsize=\figfxsize
\centerline{\epsffile{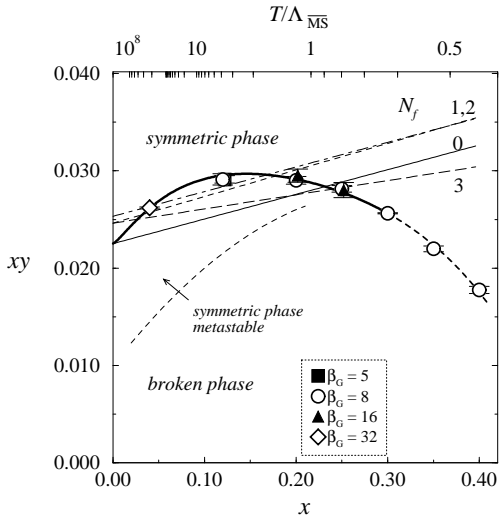}}
\figbottomspace
\caption[a]{The phase diagram of the 3D SU(2) + adjoint Higgs theory.
The plot symbols show the location of the transition given by the
simulations (extrapolated to $V\rightarrow\infty$), and
the $y=y_c(x)$ curve is a 4th order polynomial interpolation of the
data.  Dashed line indicates where the transition becomes a cross-over.
The straight lines are the `dimensional reduction lines' $y_\dr(x)$
for the number of fermion flavors indicated.} \la{fig:phasediag}
\vspace*{-1mm}
\end{figure}

The phase diagram of the SU(2) + adjoint Higgs theory is shown in
Fig.\,\ref{fig:phasediag} (for convenience, plotted in the
($x,xy$)-plane).  For SU(3) the diagram is qualitatively similar.  At
small $x$, the transition is very strongly first order, but becomes
rapidly weaker when $x$ increases.  At $x\approx 0.32$ there is a
critical point, after which only a cross-over remains.  The $N_f=0$
$y_\dr(x)$ -line is given in \eq\nr{y2}.  The 3D theory is well
defined on the entire $(x,y)$-plane, but only along $y_\dr(x)$ does the 3D
theory describe the physics of the 4D SU(2) gauge theory.  Along this
line $x$ is related to the temperature as shown on the top axis of the
plot (for SU(2), $\lambdamsbar \approx 1.2 T_c$).

In the physically relevant region $T \gg T_c \sim \lambdamsbar$ the
$y_\dr$ line is in the {\em broken phase\,}.  However, the 3D theory
cannot describe 4D high-$T$ physics in the broken phase, since the
perturbative 4D$\leftrightarrow$3D connection is not valid there
\cite{adjoint2}.  Thus the symmetric phase is the physical one.
Due to the strong 1st order nature of the
transition at small $x$, the symmetric phase is strongly {\em
metastable\,} (shown with the dashed line in
Fig.~\ref{fig:phasediag}).  If initially prepared to be in the
symmetric phase, the system remains there for the duration of any
realistic Monte Carlo simulation.

\begin{figure}[tb]
\vspace{4mm}
\figtopspace
\epsfxsize=\figfxsize
\centerline{\epsffile{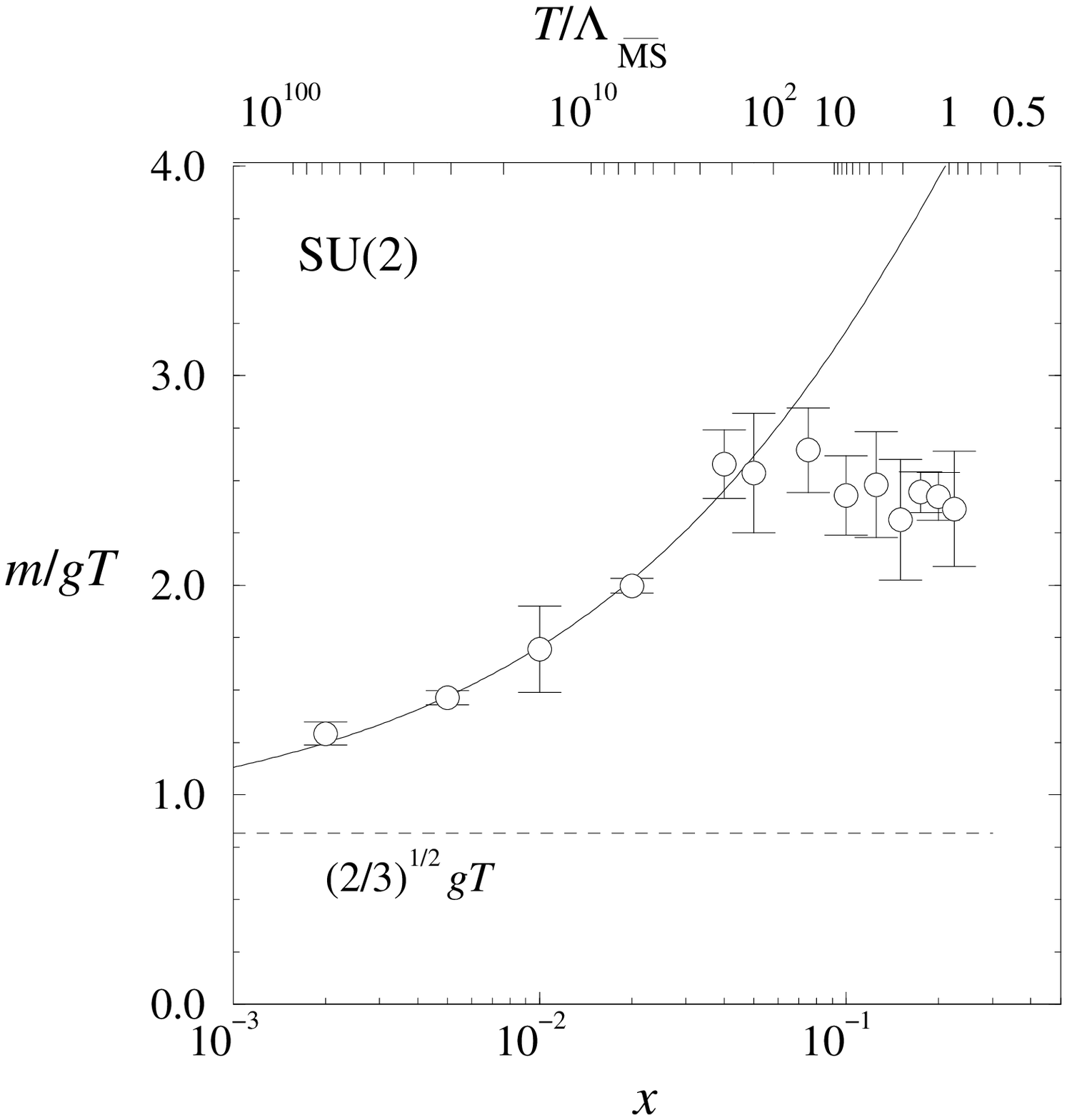}}
\vspace*{-3.6cm}
\epsfxsize=\figfxsize
\centerline{\epsffile{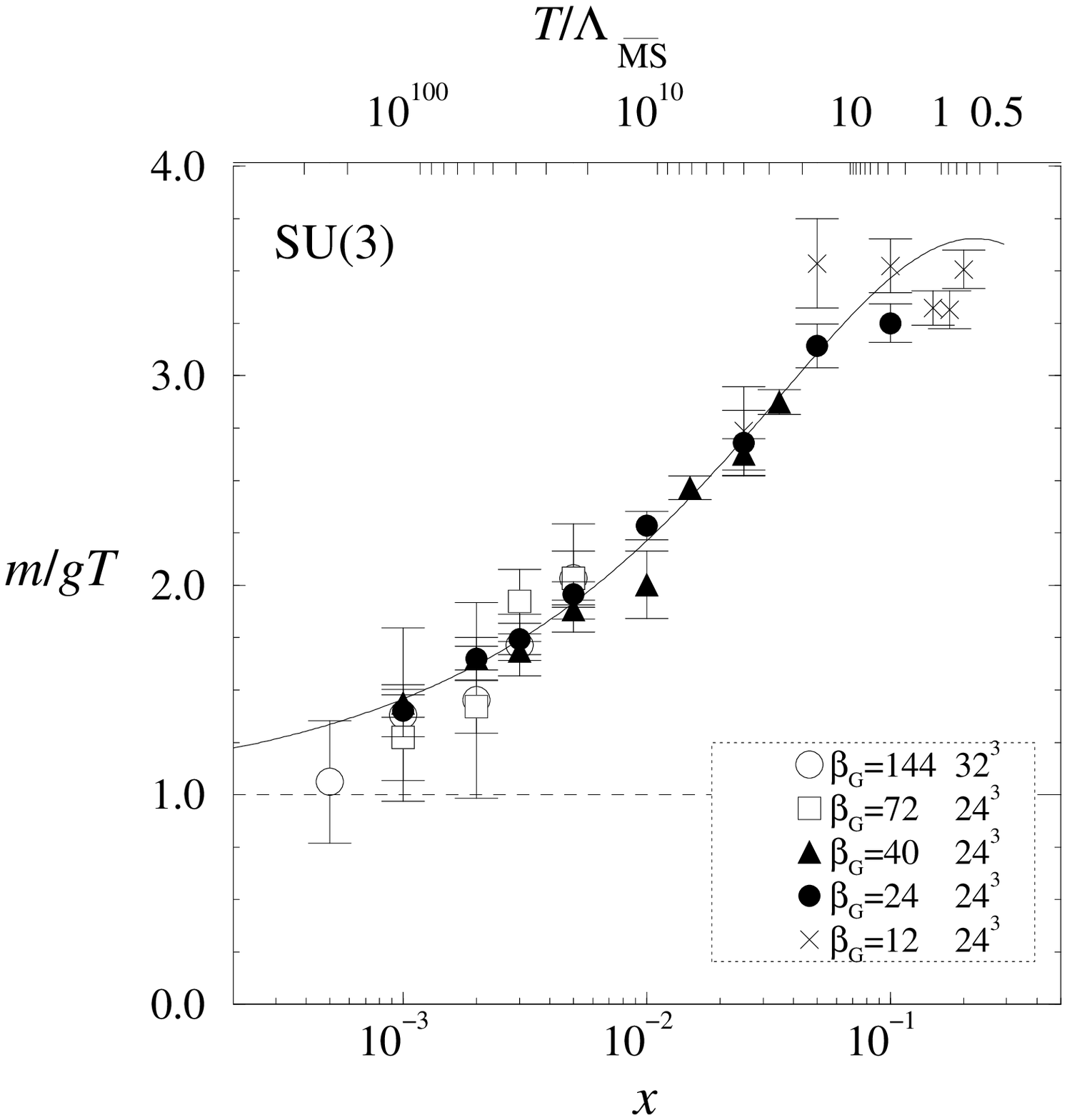}}
\figbottomspace
\caption[a]{The Debye mass 
for SU(2) (top) and SU(3) (bottom), as functions of $x$ (top scales
show $T/\lambdamsbar$ through eq.~\nr{x}).
The dashed line indicates the leading order $m_D^\rmi{LO}$, and the continuous
line the 2-parameter fit to eq.~\nr{md4d} with the parameters as in 
eq.~\nr{parameters}.}
\la{masses}
\vspace*{-4mm}
\end{figure}

The Debye mass can be defined as the mass of the lightest 3d state odd
under the reflection $A_0\rightarrow-A_0$ \cite{ay}.  
The lowest-dimensional gauge invariant operator fulfilling this is
\be
 h_i=\epsilon_{ijk}\tr A_0F_{jk}\,.
\la{hop} 
\ee
To enhance the signal, we measure the correlation function $\langle
h_i(0)h_i(z)\rangle$ using several recursive blocking levels
\cite{adjoint2}.  We perform the measurements along the metastable
$y_\dr(x)$ -lines, eqs.~(\ref{y2},\ref{y3}).  The results are shown in
Fig.~\ref{masses}, in units of 4D $gT$ ($ = g_3^2\sqrt{3y/N}$ in 3D
units).  The Monte Carlo runs are performed with several lattice
spacings $a$ ($\beta_G$).   The top scales
in Fig.~\ref{masses} show the physical temperature $T/\lambdamsbar$
along $y_\dr(x)$ -lines.  Note that the highest temperatures are
larger than $10^{100}\times \lambdamsbar \sim 10^{100} \times T_c$.

We fit the data to the 2-parameter ansatz in \eq\nr{md4d}.  At small $x$
(large $y$), the quality of the fits is very good.  We use only data
in the range $x<0.08$, so that the horizontal plateaus at large $x$
are excluded from the fits.  The results of the fits are \cite{letter}
\be\begin{array}{cll}
\mbox{SU(2):} & c_2 = 1.58(20)\,\, & b_2 =-0.03(25) \\
\mbox{SU(3):} & c_3 = 2.46(15)\,\, & b_3 =-0.49(15) \,,
\end{array}
\la{parameters}
\ee
where $b_N = d_{N,N_f}\sqrt{N/3+N_f/6}$.  For $N=2$ we can
only verify that $d_{2,N_f}$ is close to zero.  Note that writing
$c_N=N\tilde c_N$, one has $\tilde c_N =0.79\pm 0.10$ ($N=2$),
$0.82\pm 0.05$ ($N=3$).

The leading contribution to $m_D$ is dominant only at extremely large
$T$ -- indeed, for SU(3) the leading term is larger than the ${\cal
O}(g^2)$ correction for $T/\lambdamsbar\gsim 10^{19}$, implying that
the leading term only dominates when QCD anyway merges into a unified
theory.  For temperatures around $T \sim 1000\lambdamsbar$ the
non-perturbative $m_D$ is already about $3\times m_D^{\rm LO}$.  For a
detailed discussions, we refer to \cite{letter}.  Large corrections to
leading order $m_D$ have also been observed in 4D $N_f=0$ SU(2)
simulations, using gluon mass measurements in the Landau gauge
\cite{karsch}.  It remains to be seen whether this modification of the
standard picture of high-temperature gauge theories has applications
in the cosmological discussion of the quark-hadron phase transition or
in the phenomenology of heavy ion collisions.

\end{document}